 \documentclass[%
   aip,
   amsmath,amssymb,
    reprint,%
  ]{revtex4-1}
\usepackage{dcolumn}
\usepackage{bm}

\usepackage[utf8]{inputenc}
\usepackage[T1]{fontenc}
\usepackage{mathptmx}

\usepackage{color} 
\usepackage{xcolor}
\usepackage{graphicx} 
\usepackage{epsfig}

\newcommand{\bee}{\begin{equation}}
\newcommand{\ene}{\end{equation}}
\newcommand{\beea}{\begin{eqnarray}}
\newcommand{\enea}{\end{eqnarray}}

\usepackage{hyperref}
\hypersetup{colorlinks,linkcolor={blue},citecolor={blue},urlcolor={red}}  
\begin{document}
\title{Excitation and breaking of relativistic electron beam driven longitudinal electron-ion modes in a cold 
plasma}
\author{Ratan Kumar Bera}
\altaffiliation[Present address: Kevin T. Crofton Department of Aerospace and Ocean Engineering, Virginia Tech,
Blacksburg, VA 24060, USA]{}
\email{rataniitb@gmail.com}

\author{Arghya Mukherjee}
\affiliation{ 
Institute for Plasma Research, Gandhinagar-382428, Gujarat, India.%
}%

\author{Sudip Sengupta}
\affiliation{ 
Institute for Plasma Research, Gandhinagar-382428, Gujarat, India.%
}%
\author{Amita Das}
\affiliation{%
Physics Department,
Indian Institute of Technology Delhi, Hauz Khas, New Delhi-110016, India. 
}%

\date{\today}

\date{\today}
\begin{abstract}  
The excitation and breaking of relativistically intense electron-ion modes
in a cold plasma is studied
using 1D-fluid simulation techniques. To excite the mode, we have used a relativistic rigid homogeneous
electron beam propagating inside a plasma with a velocity close to the speed of light. 
It is observed that the wake wave excited by the electron
beam is
identical to the corresponding
Khachatryan mode [{\it Phys. Rev. E, 58, 6(1998)}], a relativistic electron-ion mode in a cold plasma. 
It is also seen in the simulation that
the numerical profile of the excited electron-ion mode gradually modifies
with time and eventually breaks after several plasma periods exhibiting explosive behavior in 
the density profile. This is an well known phenomena, known as ``wave breaking". It is found that
the numerical wave breaking limit of these modes lies much below than their
analytical breaking limit. The discrepancy between the numerical and analytical wave breaking limit has been understood 
in 
terms of phase-mixing process of the mode. The phase 
mixing time (or wave breaking time)
obtained from the simulations has also been scaled as a function of beam parameters
and found to follow the analytical scaling.
\end{abstract}

\maketitle 
\section{Introduction}
Over decades, the research of relativistically strong plasma waves (RSWs) has attracted
significant amount of attention because of its active contribution to the progress of 
plasma physics as well as astrophysics. For instance the excitation and breaking of such
waves serves a useful paradigm to illustrate the physics of plasma based 
accelerations schemes \cite{tajima,chen,esarey,uhm,ratan}, 
fast ignition concept in the Inertial confinement fusion systems \cite{tabak,Pukhov,Wills}, as well as
various solar and astrophysical interdisciplinary
processes \cite{Ferrari,ebisuzaki,shukla,Tsiklauri}. Based on the
ion dynamics, the relativistic propagating modes in a
cold unmagnetized plasma can be typically categorized into two types,
``Akhiezer-Polovin mode'' \cite{akhiezer}
and ``Khachtryan mode'' \cite{khacha}. 
Akhiezer Polovin mode is basically a relativistic electron plasma wave 
when the effect of ion 
motion are completely ignored mostly due to their heavy mass. On the other hand Khachtryan 
mode refers to a relativistic electron-ion mode when both the dynamics of
electrons as well as 
ions
are important. In general, every RSWs
in a cold plasma are nothing but the Khachtryan modes.
Akhiezer-Polovin (AP) modes are such Khachtryan modes
at which the ions are immobile.
Though the contribution of the ion motion on the collective dynamics of plasmas are 
negligibly small in most of the laboratory and astrophysical events, but their effects are
still crucial in various phenomena. 
For 
example, in most of the astrophysical systems, mostly around the pulsars, it 
is considered to be filled with electron-positron plasmas, namely pair plasmas
(plasmas consisting of two classes of particles with opposite sign of the charge, but equal mass) \cite{Arons,Hirotani}.
In such scenarios, the excitation of relativistic electron-ion modes and their breaking 
are critical for understanding many astrophysical events like jet formation,
Ultra-High-Cosmic-Rays (UHECRs) generation, shock acceleration process etc. \cite{Ferrari,Hirotani,ebisuzaki}.
Furthermore,
Khachatryan et al. \cite{khacha} reported on the study of strong plasma waves in presence of ions that 
plasma ions (even for heavy ions) can make an essential contribution to the process of charge 
separation in relativistic regimes.
The recent reports on plasma based particle acceleration process where RSWs are 
used to accelerate the charge particles into high energies indicate that 
the motion of ions can give rise 
to the transverse fields that in turn disrupt the motion of the driver beam propagation and they also can effect
the energy transfer 
ratio from the driver beam
to the accelerated particles \cite{rosen2,rosenion,viera}. Therefore, the excitation and breaking of 
relativistic electron-ion modes is still an active area of research.
To date, the excitation of relativistic electron modes (Akhiezer-Polovin (AP) modes)
and their breaking have been well examined by 
several authors \cite{prabal,arghya,ratan2}. It has been well established 
that AP mode can break much below its wave breaking limit due to phase mixing process if
it is subjected to a arbitrary amplitude of longitudinal perturbation.
The analytical scaling of phase mixing time for AP mode has also been verified with 
the numerical works \cite{prabal,arghya,ratan2}. 
Such an extensive study for the excitation and breaking of 
relativistic electron-ion mode or Khachatryan mode is still an unexplored area of research.

\par\vspace{\baselineskip} 
In this paper, we have investigated thoroughly the excitation and breaking
of relativistic longitudinal electron-ion modes or Khachatryan modes in a cold plasma
using 1-D fluid simulation techniques. To excite the mode,
we have used a rigid homogeneous
pulsed electron beam which propagates
inside a clod unmagnetized homogeneous plasma with a speed
close to the speed of light. 
Here both the dynamics of plasma electrons and ions are considered for exciting the relativistic electron-ion modes.
When an electron beam propels inside the plasma it expels the nearby plasma electrons and 
attracts the ions (in case of mobile ions)
due to the space-charge force. Typically the beam displaces the plasma electron and ions in opposite direction. 
In case of immobile ions, the displacement of ions will be negligibly small. They 
will only provide a static neutralizing background.
However, as the beam propagates further inside the plasma, the displaced plasma electrons as well as
ions will try to come back to their 
original position to nullify the charge separation. But due to their inertia, they will overshoot their original 
position.
As a result an oscillation or a wave will be established at the wake of the beam
having phase velocity equal to the velocity of the beam. 
In other words, a relativistic electron-ion mode will be excited at the wake of the beam propagating
with a phase velocity equal to the velocity of the beam.In the present paper,
With the help of three fluid description of the beam plasma medium, the excitation and spatio-temporal evolution of relativistic electron beam driven wake waves or relativistic electron-ion modes
in a cold plasma has been studied using 1-D fluid simulation techniques.
It is shown that
the numerical profiles of the wake wave obtained from the fluid simulation
show a good agreement with their corresponding analytical
profiles given by Rosenzweig et al.\cite{rosenion} for different beam density.
It is also observed that the wake wave excited by the $e^{-}$ beam is identical to a corresponding
Khachtryan's mode \cite{khacha}. 
Furthermore, it is seen during the space-time evolution of the excited wave in the simulation that the profile of the excited wave gradually modifies with time and
deviates significantly from the analytical solution of Khachtryan mode as well as Rosenzweig's solution.
After several plasma periods, we see that the density profile associated with the wake wave becomes spiky which is a clear signature of wave breaking \cite{ratan2,arghya,sudip_pre,prabal2,infeld,sudip_ppcf,arghya3}. When the wave breaks or density bursts form in the simulation,
we have calculated the maximum amplitude of the electric field or wave breaking limit of the wave.
It is observed that the numerical wave breaking limit
lies much below than the analytical limit given by Khachatryan et al. \cite{khacha}. 
This inconsistency of the analytical and numerical wave breaking limit has been understood in terms of phase mixing 
process \cite{ratan2,prabal,sudip_pre,arghya}. 
It has been shown that the electron-ion mode breaks much below its analytical wave breaking limit 
due to the gradual process of phase mixing. From the simulation, 
we have obtained the phase mixing time or wave breaking time and plotted it with respect to the wave parameters.
It is found that the numerical curve follows the analytical scaling given by Arghya et al. \cite{arghya3}.
\par\vspace{\baselineskip}
In next section (Section -II),
we present the basic equations governing the 
excitation of relativistic electron-ion mode driven by relativistic electron beam in a cold plasma. 
We have discussed our numerical techniques for this study in section-III. 
Our numerical observations and a detail discussion of these results have been covered in section-IV. 
A brief summary of our work presented in this paper is covered in section V.
\par\vspace{\baselineskip}
\section{Governing Equations}
%
The basic equations governing the excitation of longitudinal relativistic
electron ion mode driven by electron beam
in a cold plasma are the relativistic fluid-Maxwell equations. These equations contain the continuity
and momentum equations for plasma electrons, plasma ions, and also for electron beam. The Poisson's
equation have been used to calculate the electric field in the system.
Therefore, the basic normalized
governing equations for the study of longitudinal ($z$-direction) electron-ion modes in plasmas are,    
\begin{equation}
  \frac{\partial n_e}{\partial t}+\frac{\partial (n_e v_e) }{\partial z}=0	\label{pl_cont}
 \end{equation}
 \begin{equation}
  \frac{\partial p_e }{\partial t}+v_e\frac{\partial p_e}{\partial z}=-E      \label{pl_mom}
 \end{equation}

 \begin{equation}
  \frac{\partial n_i}{\partial t}+\frac{\partial (n_i v_i) }{\partial z}=0	\label{ion_cont}
 \end{equation}
 \begin{equation}
  \frac{\partial p_i }{\partial t}+v_i \frac{\partial p_i}{\partial z}=-\mu  E     \label{ion_mom}
 \end{equation}

  \begin{equation}
  \frac{\partial n_b}{\partial t}+\frac{\partial (n_b v_{b}) }{\partial z}=0  \label{beam_cont}
 \end{equation}
 \begin{equation}
 \frac{\partial p_{b} }{\partial t}+v_{b}\frac{\partial p_{b}}{\partial z}=-E \label{beam_mom}
 \end{equation}
 \begin{equation}
\frac{\partial E }{\partial z}=(n_i-n_e-n_b) \label{pois}
 \end{equation}

 where $p_e=\gamma_e v_e$, $p_i=\gamma_i v_i$ and  $p_b=\gamma_b v_b$  are the $z$-components of
 momentum of plasma electron, plasma ion  and beam electron having $z$-component of velocity 
 $v_e$, $v_i$ and $v_b$ respectively. Here, $\gamma_e=\left(1-v_e^2\right)^{-1/2}$, $\gamma_i=\left(1-v_i^2\right)^{-1/2}$ 
 and $\gamma_b=\left(1-v_b^2\right)^{-1/2}$ are the relativistic 
factors associated with plasma electron, plasma ion and beam electron respectively.  
In the above equations, $n_e$, $n_i$ and $n_b$ represents the density of plasma electron, plasma ion and electron beam 
respectively.  $E$ and $\mu=m_e/m_i$ represents the longitudinal electric field and mass ratio 
(ratio of electron to ion mass) 
respectively. We have used the normalization factors as,
$t \rightarrow \omega_{pe}t$, $z \rightarrow \frac{\omega_{pe}z}{c}$, $E \rightarrow \frac{eE}{m_e c\omega_{pe}}$,
$v_e\rightarrow \frac{v_e}{c}$, $v_i\rightarrow \frac{v_i}{c}$, $v_b\rightarrow \frac{v_b}{c}$
$p_e \rightarrow \frac{p_e}{m_e c}$, $p_i \rightarrow \frac{p_i}{m_e c}$, $p_b \rightarrow \frac{p_b}{m_e c}$,
$n_e\rightarrow \frac{n_e}{n_0}$, $n_i\rightarrow \frac{n_i}{n_0}$ and $n_b\rightarrow \frac{n_b}{n_0}$.
Here $\omega_{pe}$, $n_0$, and $c$ are the electron plasma frequency, equilibrium plasma density, and speed 
of the light respectively. 
\par\vspace{\baselineskip}
In this paper, we have used a rigid beam to excite the waves of
constant amplitude and phase velocity \cite{ratan2,rosenzweig}. Ideally a rigid beam defines such a beam which
can penetrates infinite length inside plasma without any deformation.
This is true only for a sufficiently energetic beam.
In our earlier works \cite{ratan,ratan2}, we have shown that  the beam can be considered to be rigid 
for hundred of plasma periods only if the velocity of the beam $v_b \geq 0.99$.
In this limit, the evolution equation(\ref{beam_mom}) for the beam can be ignored
and the propagation of the rigid beam can be completely depicted by the equation (\ref{beam_cont}).
Therefore, we finally have 
equations (\ref{pl_cont}), (\ref{pl_mom}), (\ref{ion_cont}), (\ref{ion_mom}), (\ref{beam_cont}), and (\ref{pois})
which are the key equations to study the excitation and breaking 
of relativistic longitudinal electron-ion mode driven by relativistic electron beam in a cold plasma.
Below we discuss the numerical techniques used to solve these equations
((\ref{pl_cont}), (\ref{pl_mom}), 
(\ref{ion_cont}), (\ref{ion_mom}), (\ref{beam_cont}) and (\ref{pois})).

\section{Fluid simulation techniques}
In this section, we present the numerical techniques used to study the excitation and breaking of 
relativistic electron-ion mode
in a cold plasma.
We have developed a 1-D fluid code using LCPFCT subroutines based on flux-corrected transport (FCT) scheme \cite{boris}. 
The 
basic principle of FCT scheme is based on the generalization of two-step Lax-Wendroff method \cite{numr}.
LCPFCT subroutines are mainly used to solve generalized continuity like equations ((\ref{pl_cont}), (\ref{pl_mom}), 
(\ref{ion_cont}), (\ref{ion_mom}), (\ref{beam_cont})). To solve the Poisson equation (\ref{pois}),
we have used successive over-relaxation (SOR) method. 
Coupling these schemes iteratively, we have developed the 1-D fluid code and solved
the equations (\ref{pl_cont}), (\ref{pl_mom}), (\ref{ion_cont}), (\ref{ion_mom}), (\ref{beam_cont})
and (\ref{pois}) numerically. In the simulation,
the driver beam is allowed to propagate from one end to the 
other end of the simulation box along $z$-direction.
For a given beam profile,
we have initiated the simulation using the corresponding analytical profiles of 
plasma electron density, ion density, electron velocity, ion velocity, and electric field
given by Rosenzweig et al. (\cite{rosenion}) and then followed the space-time evolution of the system. 
The results obtained from simulation are checked repeating the simulations for different mesh sizes. 
The code has also
bench-marked studying various standard problems \cite{ratan, ratan2}.
In next section, we discuss the simulation results in detail.
\par\vspace{\baselineskip}
\section{Results and discussion}
\subsection{Excitation of the relativistic electron-ion modes in a cold plasma using electron beam}
\begin{figure*}
    \includegraphics[width=0.9\textwidth]{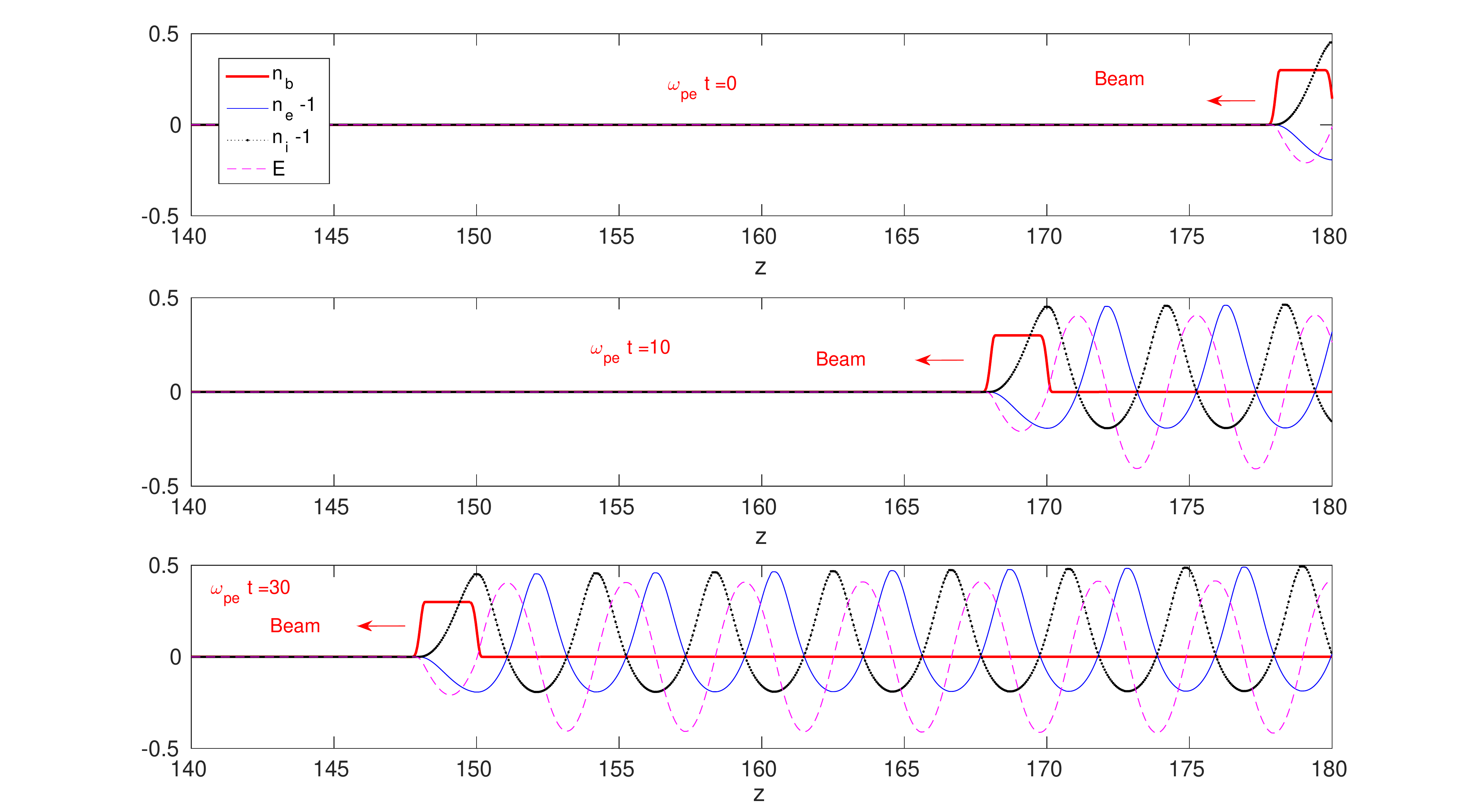}
    \caption{Plot of normalized perturbed electron density ($n_e -1$),  ion density  ($n_e -1$), and electric field ($E$)
    profiles at different times for the normalized beam density ($n_b$)=0.3, beam length ($l_b$)=4, beam velocity ($v_b$)
    =0.9999 and mass ratio ($\mu$) =1.}
    \label{fig1}
\end{figure*}

\begin{figure*}
    \includegraphics[width=0.9\textwidth]{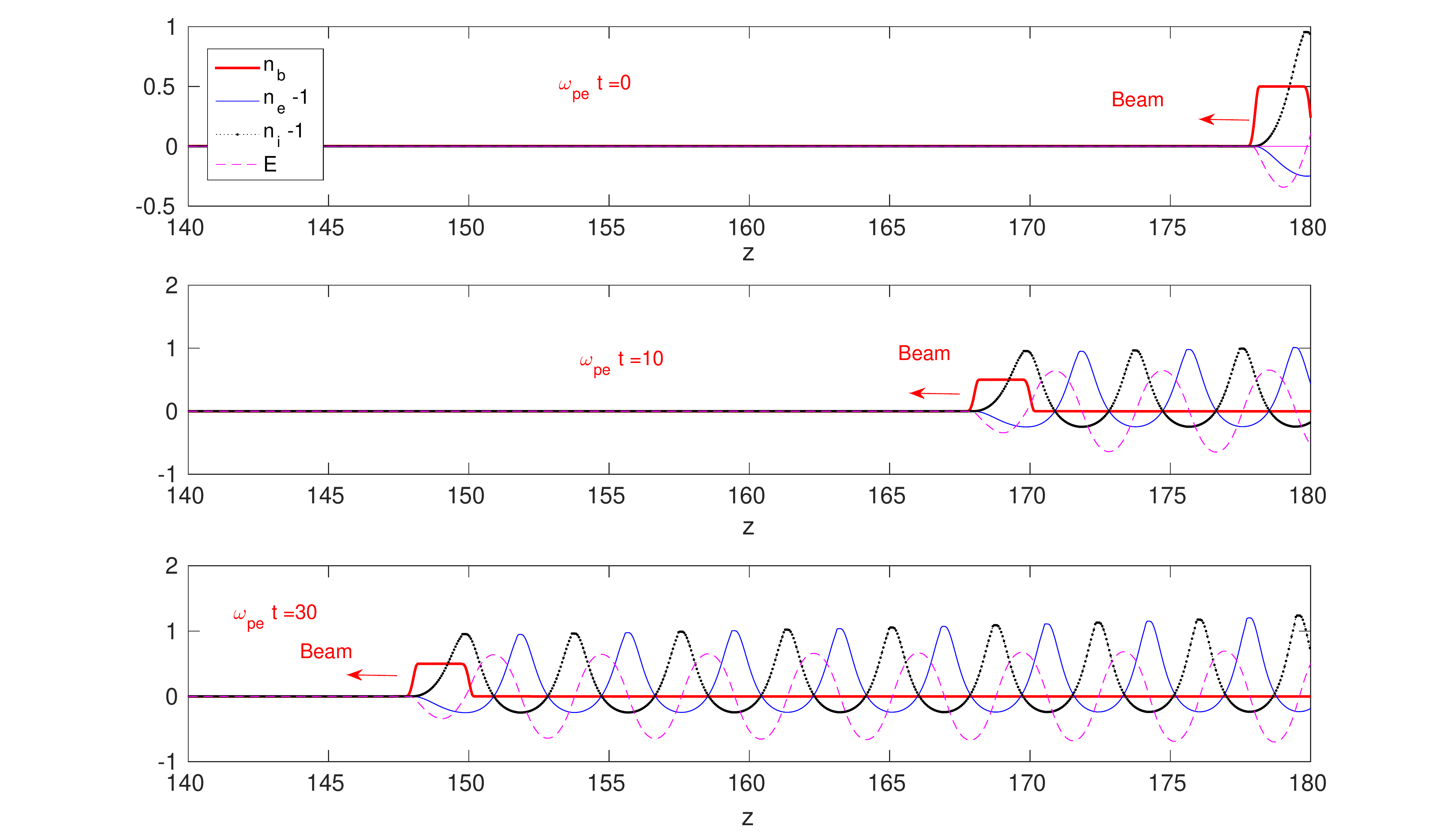}
           \caption{Plot of normalized perturbed electron density ($n_e -1$),  ion density  ($n_e -1$), and electric field ($E$)
    profiles at different times for the normalized beam density ($n_b$)=0.5, beam length ($l_b$)=4, beam velocity ($v_b$)
    =0.9999 and mass ratio ($\mu$) =1.}
       \label{fig2}
\end{figure*}

\begin{figure*}
    \includegraphics[width=0.9\textwidth]{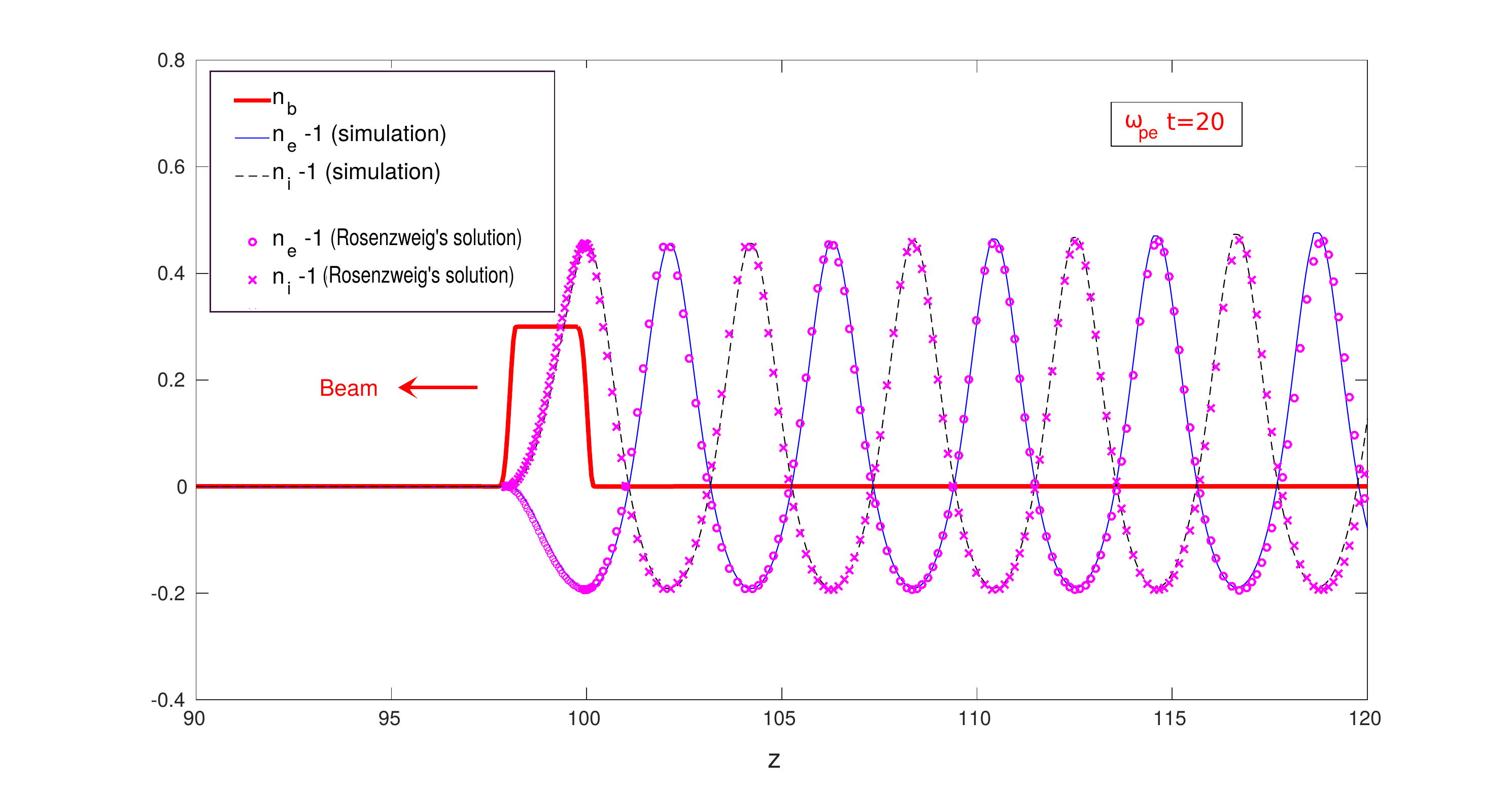}
      \caption{Plot of numerical and analytical normalized perturbed electron density ($n_e -1$) and ion density ($n_i -1$)
    profiles at $\omega_{pe} t =20$ for the normalized beam density ($n_b$)=0.3, beam length ($l_b$)=4,
    beam velocity ($v_b$) =0.9999 and mass ratio ($\mu$) =1.}
   
    \label{fig3}
\end{figure*}

Here we present the simulation results for the excitation of relativistic electron ion-modes
in a cold plasma using electron beam. 
The simulations have been performed for different beam densities ($n_b$) and mass ratios ($\mu$).
In all the simulations, we have kept fixed the beam velocity $v_b=0.9999$ and beam length $l_b=4$. 
Hence the phase velocity ($v_{ph}$) of the excited wake wave is fixed to $0.9999$.
The amplitude and the frequency of excitation is now completely determined by the values of $n_b$ and $\mu$.
By changing the values of $n_b$ and $\mu$ one can excite wake waves of different amplitude and frequency.
In Figs. (\ref{fig1}) and (\ref{fig2}), we have plotted
the perturbed plasma electron density ($n_e -1$), ion density ($n_e -1$), and electric field ($E$)
profiles at different times $\omega_{pe}t =0, 10$, and $30$ for $n_b$=0.3 and $n_b$=0.5 respectively; where $\mu$ =1. 
To initiate the simulations at $\omega_{pe}t=0$ in each cases, 
we have used the analytical profiles of $n_e-1$, $n_i-1$, $v_e$, $v_i$, and $E$ by
solving equations (20-23) in ref. \cite{rosenion}. We see that the beam excites wake 
wave as it passes through the plasma.
It is seen that the amplitude of the wake wave increases by increasing $n_b$ for a given value of $\mu$.
For the sake of completeness we
have next plotted the analytical solutions ($viz$ Rosenzweig's solution) of beam driven wake wave by
solving equations (20-23) from ref. \cite{rosenion}
on top of the corresponding numerical profile for $n_b=0.3$ in Fig. (\ref{fig3}).
We see that the simulation results shows a good agreement with the analytical results. 
\par
\begin{figure*}
    \includegraphics[width=0.9\textwidth]{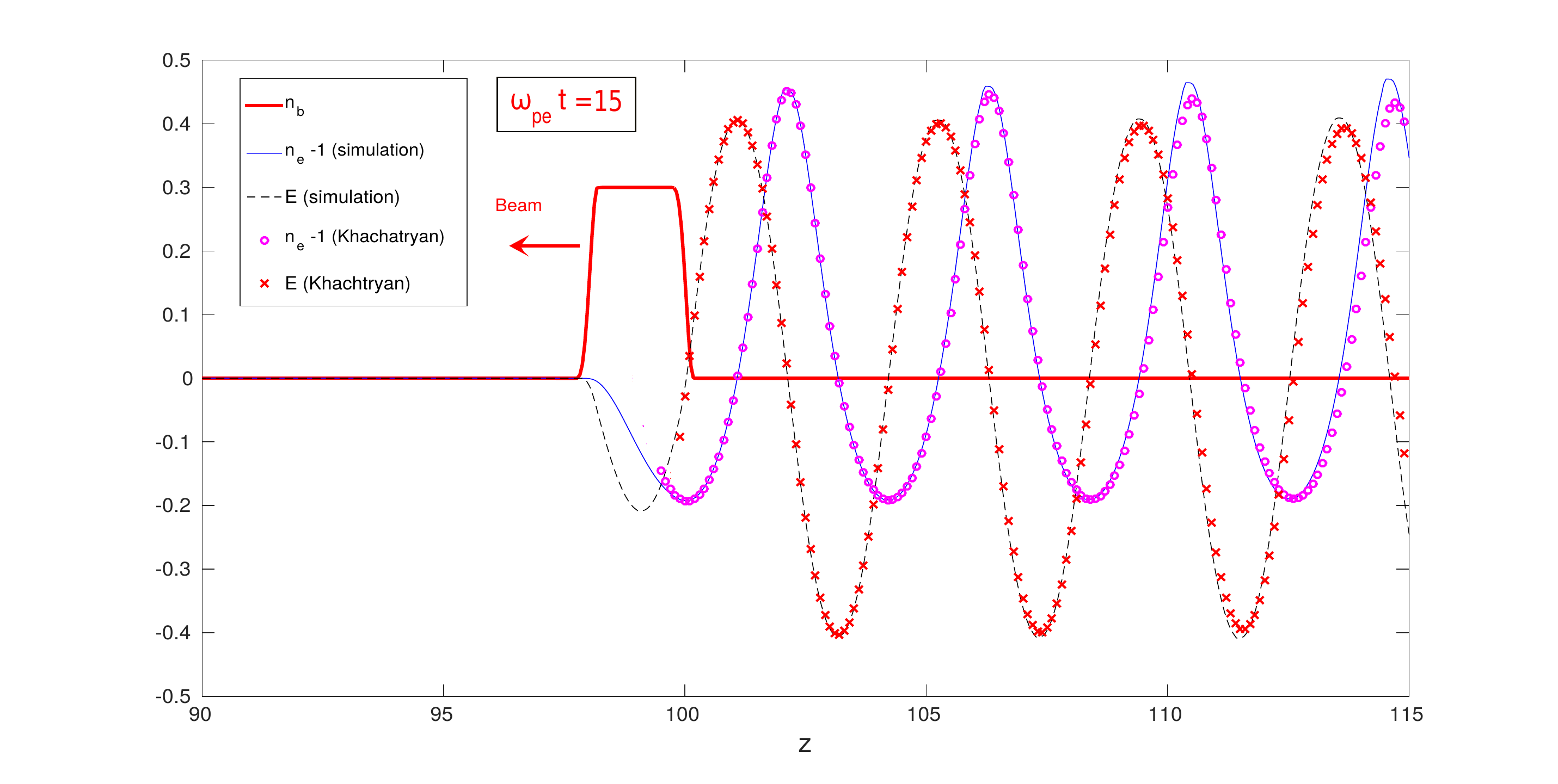}
      \caption{Plot of normalized perturbed electron density ($n_e -1$) and electric field ($E$)
    profiles obtained from simulation and corresponding Khachatrayan mode (analytical) 
    at $\omega_{pe} t =15$ for the normalized beam density ($n_b$)=0.3, beam length ($l_b$)=4,
    beam velocity ($v_b$) =0.9999 and mass ratio ($\mu$) =1.}
       \label{fig4}
\end{figure*}

\begin{figure*}
    \includegraphics[width=0.9\textwidth]{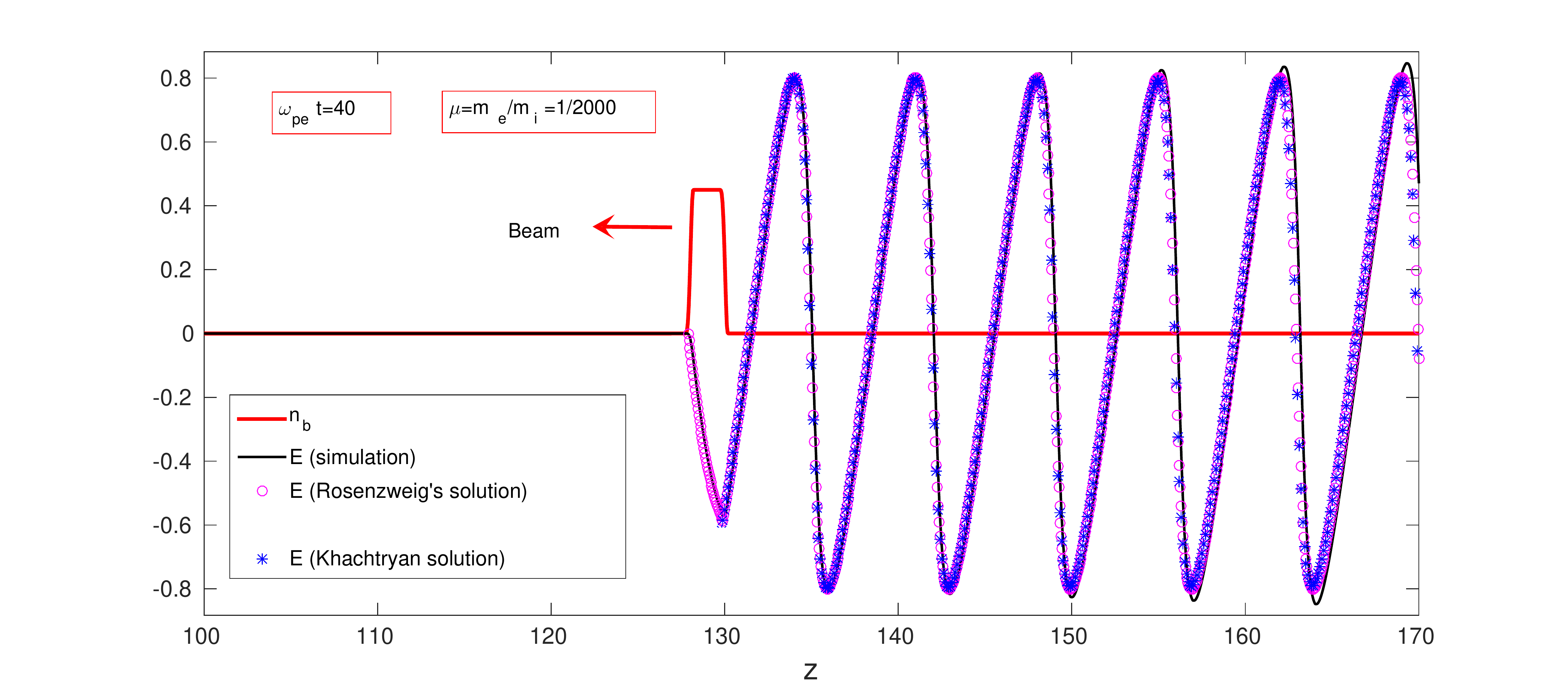}
    \caption{Plot of numerical and analytical profiles of normalized electric field ($E$) 
    for the mass ratio ($\mu$)=$1/2000$ at $\omega_{pe} t =40$ for the normalized beam density ($n_b$)=0.5, beam length ($l_b$)=4,
    beam velocity ($v_b$) =0.9999.}
    \label{fig5}
\end{figure*}

\vspace{\baselineskip}
Furthermore,it is to be noted that the equations (\ref{pl_cont}), (\ref{pl_mom}), (\ref{ion_cont}), (\ref{ion_mom}), 
and (\ref{pois}) with the term $n_b=0$ are nothing but the relativistic fluid-Maxwell equations in 1-D. The wave-frame
solutions of these equations
are the well known solutions of Khachatrayan mode (see sec II in ref. \cite{khacha}) which is
a relativistic electron-ion mode in a cold plasma. The beam density vanishes at the wake of the beam i.e. $n_b =0$.
Therefore we expect that 
the wake wave excited by the beam must be a corresponding Khachatryan mode.
A Khachtryan mode can be parameterized in terms of $E_{max}$, $v_{ph}$, and $\mu$; where $E_{max}$ 
represents the maximum amplitude of the electric field of the excitation. For a given values of
$E_{max}$, $v_{ph}$, and $\mu$,
one can easily show the corresponding structure of a 
Khachatryan mode by solving equation (4-9) of ref. \cite{khacha}.
Using the numerical values of $E_{max}$, $v_{ph}=v_b$, and $\mu$ of the excited wake wave from 
the simulation for $n_b=0.3$, 
we have solved equations (4-9) of ref. \cite{khacha} and plotted the corresponding Khachatryan mode
on top of the wake wave in Fig. (\ref{fig4}). 
It is seen that the wake wave is nothing but a corresponding Khachatryan mode propagating with a phase velocity 
equal to the velocity of the beam. 
To further emphasize the fact that the wake wave excited by the relativistic electron beam is a
corresponding Khachatryan mode, we have plotted for different values of $n_b=0.5$ and also 
$\mu=1/2000$ in Fig. (\ref{fig5}). We see that Khachatrayan mode converts to Akhiezer-Polovin mode for low $\mu$.
\subsection{Breaking of relativistic electron-ion modes in a cold plasma}
\begin{figure*}
    \includegraphics[width=0.9\textwidth]{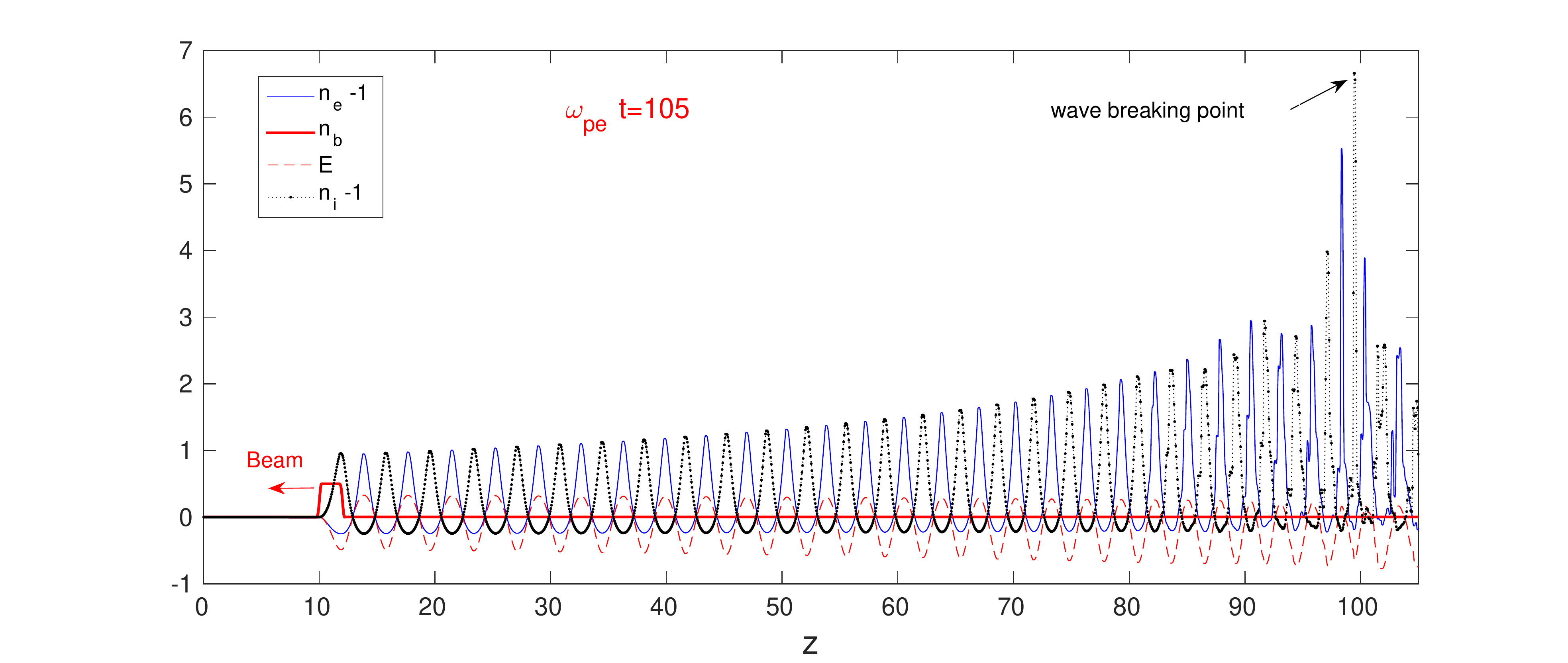}
  \caption{Plot of numerical profile of normalized perturbed electron density ($n_e -1$),  ion density  ($n_e -1$), and electric field ($E$)
   at $\omega_{pe} t =125$ for the normalized beam density ($n_b$)=0.3, beam length ($l_b$)=4, beam velocity ($v_b$)
    =0.9999 and mass ratio ($\mu$) =1.}
    \label{fig6}
\end{figure*}

\begin{figure*}
    \includegraphics[width=0.9\textwidth]{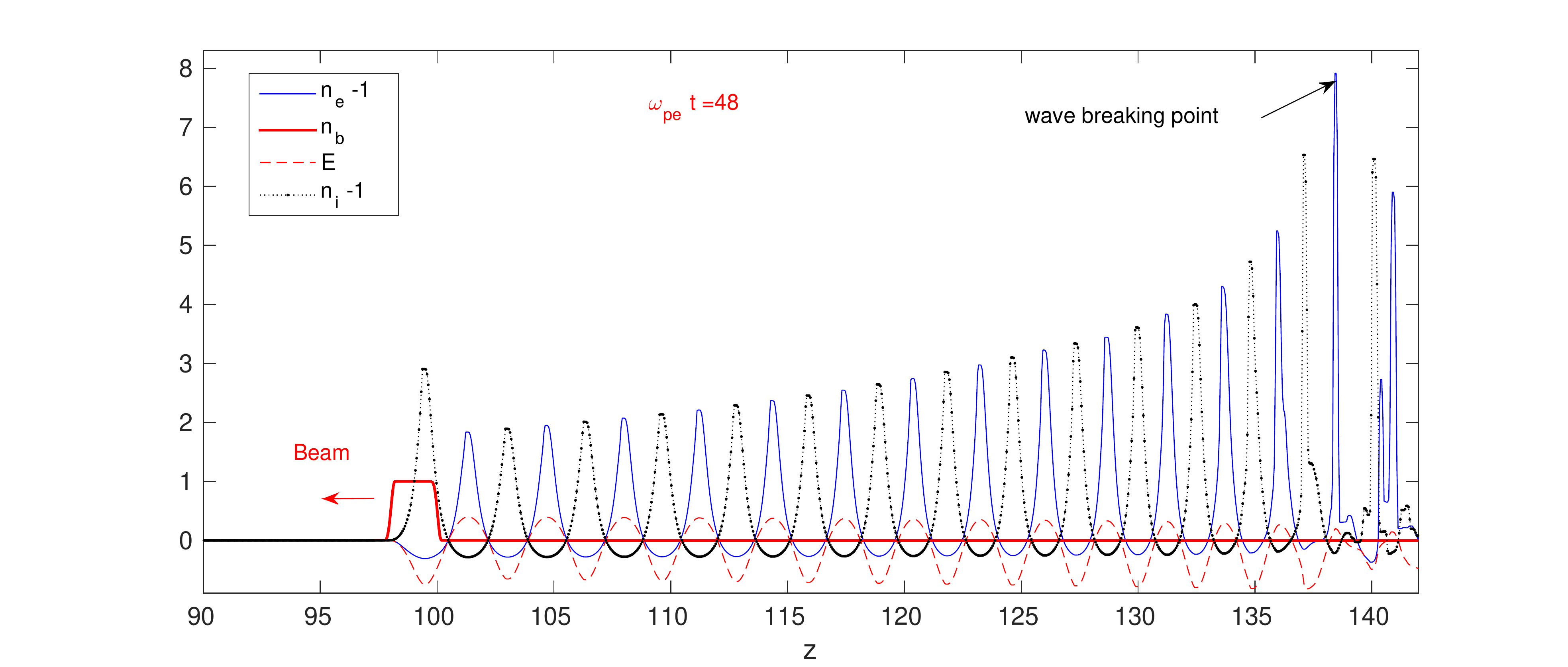}
    \caption{Plot of numerical profile of normalized perturbed electron density ($n_e -1$),  ion density  ($n_e -1$), and electric field ($E$)
   at $\omega_{pe} t =125$ for the normalized beam density ($n_b$)=1, beam length ($l_b$)=4, beam velocity ($v_b$)
    =0.9999 and mass ratio ($\mu$) =1.}
       \label{fig7}
\end{figure*}

\begin{figure*}
    \includegraphics[width=0.9\textwidth]{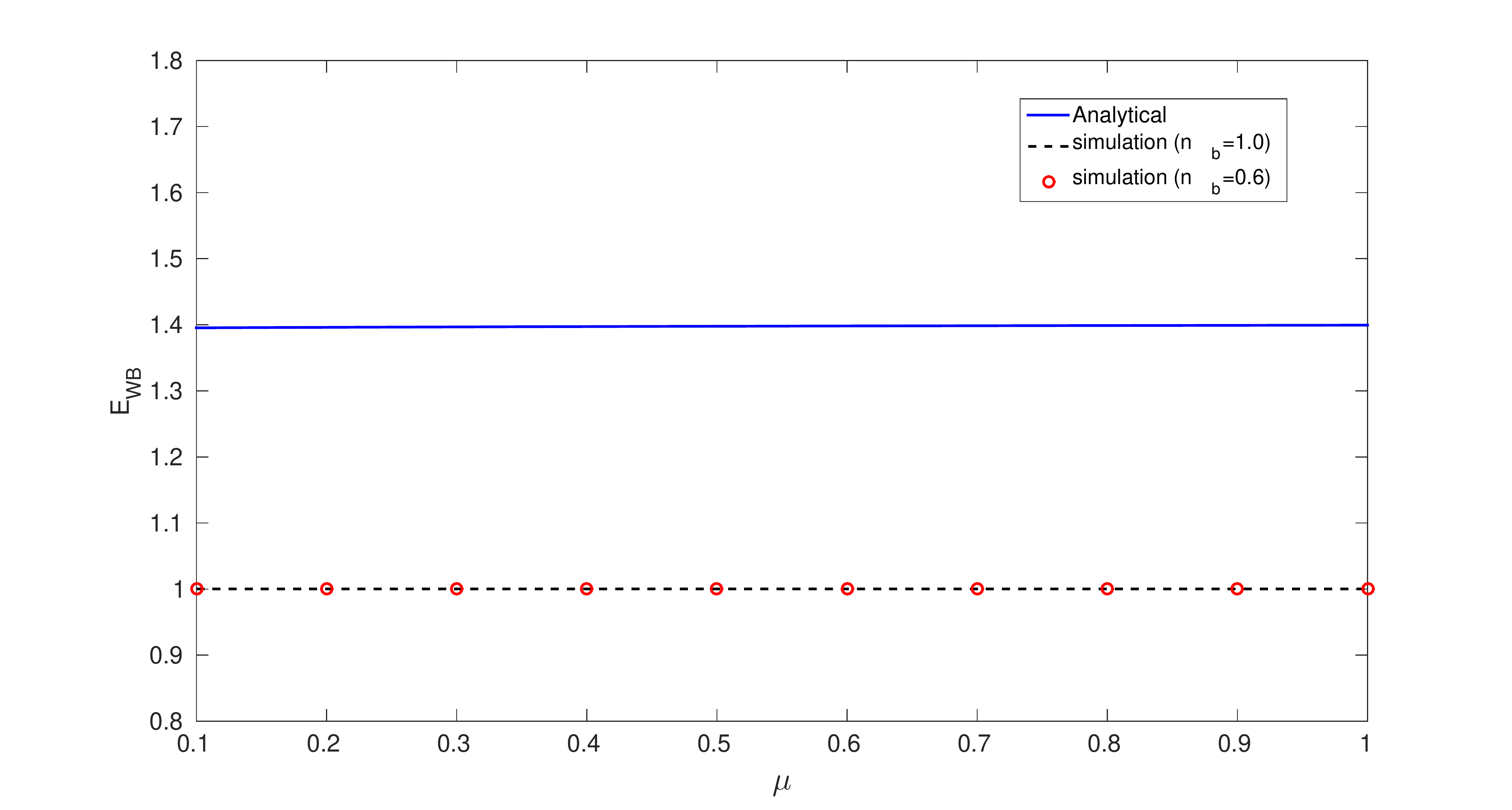}
\caption{Plot of theoretical and numerical wave breaking electric field ($E_{WB}$)  vs. mass ratios ($\mu$) 
for normalized beam density $n_b=0.6$ and $n_b=1$ ; where beam length ($l_b$)=4, beam velocity ($v_b$) =0.9999}
    \label{fig8}
\end{figure*}

Next, we have also observed several other interesting features by
following the space-time evolution of the excited wave in the simulation for a long time. 
In Figs. (\ref{fig6})
and (\ref{fig7}), we have plotted the profiles of the
perturbed electron density ($n_e -1$), ion density  ($n_e -1$), and electric field ($E$) for $n_b=0.5$ 
at $\omega_{pe}t=105$ and $n_b=1$ at $\omega_{pe}t=48$
respectively.
We see that
the numerical profiles of electron density, ion density and electric field get modified with time and
deviate significantly from their analytical 
profiles after several plasma periods. It is also seen that
the amplitude of the electron density and ion density gradually increases
with time and eventually acquire spiky structure 
at later times. After a certain time
(e.g. $\omega_{pe}t=90$ for $n_b=0.5$ (see Fig. \ref{fig6}) and $\omega_{pe}t=38$ for $n_b=1$ (see Fig. \ref{fig7})), 
the amplitude of the density profile (electron and ions) becomes maximum and
suddenly decreases afterwards. The electric field amplitude 
also gets suppressed after this critical time. This is a clear signature of wave breaking 
\cite{sudip_ppcf,sudip_pre,arghya,ratan2,arghya2}. The time 
at which the amplitude of the density spikes get maximized and then goes down is known as wave breaking time.
Typically in a medium, 
a wave breaks when the amplitude of the wave reaches to its maximum value,
known as wave breaking limit 
\cite{prabal,sudip_pre}. If the amplitude of the excitation crosses that limit,
the wave can not be sustained by the medium and the wave finally breaks.
The wave breaking limit for a plasma wave can be defined by the amplitude 
of its electric field ($E_{WB}$). For Khachatryan mode or a relativistic electron ion mode,
the analytical expression of the wave breaking limit 
is given by Khachatryan et al.\cite{khacha} as,
$E_{WB}=\sqrt{2} \gamma_{ph} [1+(1-\xi_1^\frac{1}{2} \xi_2^\frac{1}{2})/\mu]$ 
;
where $\xi_1 =1+\mu$, $\xi_2 =1+[\mu(\gamma_{ph}-1)/(\gamma_{ph}+1)]$ and $\gamma_{ph}=(1-v_{ph}^2)^{-\frac{1}{2}}$. 
When the wake wave or Khachatryan mode breaks in the simulation for $n_b=0.6$ and $n_b=1$, 
we have determined the maximum amplitude of the electric field 
for different values of $\mu$ 
and plotted in Fig. (\ref{fig8}) as a function of $\mu$ along with their corresponding analytical values. 
We have seen that 
the numerical wave breaking limit lies much below the analytical wave breaking limit.
This indicates that the numerically excited electron-ion mode 
breaks before it touches to its analytical wave breaking limit. 
\par\vspace{\baselineskip}
The deviation between the analytical and the numerical breaking limit 
of relativistic propagating electron-ion modes 
has been understood in terms of phase mixing process \cite{sudip_ppcf,arghya,ratan2}. 
Normally phase mixing process occurs
when the frequency of the wave becomes space dependent. As a result 
the neighboring fluid elements or particles sustaining the wave will oscillate with different frequencies. Hence 
the profile of the wave gets modified with time gradually . 
Eventually a time will occur when
the two neighboring elements will oscillate with out of phase. As a consequence they will cross each other and 
the wave will break exhibiting sharp spikes in the density profile at the crossing point. After the breaking, the wave will loose its coherency
and the energy of the wave goes to random particle motion.
The effect of phase mixing process on the wave-breaking phenomena of 
relativistic electron plasma waves or Akhiezer-Polovin modes
has already been  extensively studied
by several authors \cite{sudip_pre, sudip_ppcf, prabal,prabal2,arghya}. They have shown that AP mode can break 
much below its wave breaking limit due to phase mixing process.
In one of our earlier works \cite{ratan2} for immobile ions, 
we have shown that the wake wave excited by relativistic beam 
is nothing but a corresponding Akhiezer-Polovin (AP) mode \cite{akhiezer}
and breaks much below the analytical wave breaking limit.
Normally a pure Akhiezer-Polovin mode never breaks as its wave breaking limit $E_{WB} = \sqrt{2(\gamma_{ph} -1)}$.
For relativistic AP modes, $E_{WB} \rightarrow \infty$ as $\gamma_{ph} \rightarrow \infty$.
In 2007, Prabal et al. \cite{prabal} showed that AP mode can break below the wave breaking limit
if it is subjected to a longitudinal perturbation of an arbitrary amplitude. 
Due to the longitudinal perturbations, the frequency of the wave becomes space dependent. As a result,
the wave breaks via phase mixing process before it reaches to its wave breaking limit. 
In the simulation for immobile ions, the beam excites pure AP mode at its wake.
The inherent numerical fluctuation then gets superimposed with this wave and acts as a perturbation on it. 
As a results the wake wave or AP mode breaks via phase mixing process. 
\par
\begin{figure*}
    \includegraphics[width=0.9\textwidth]{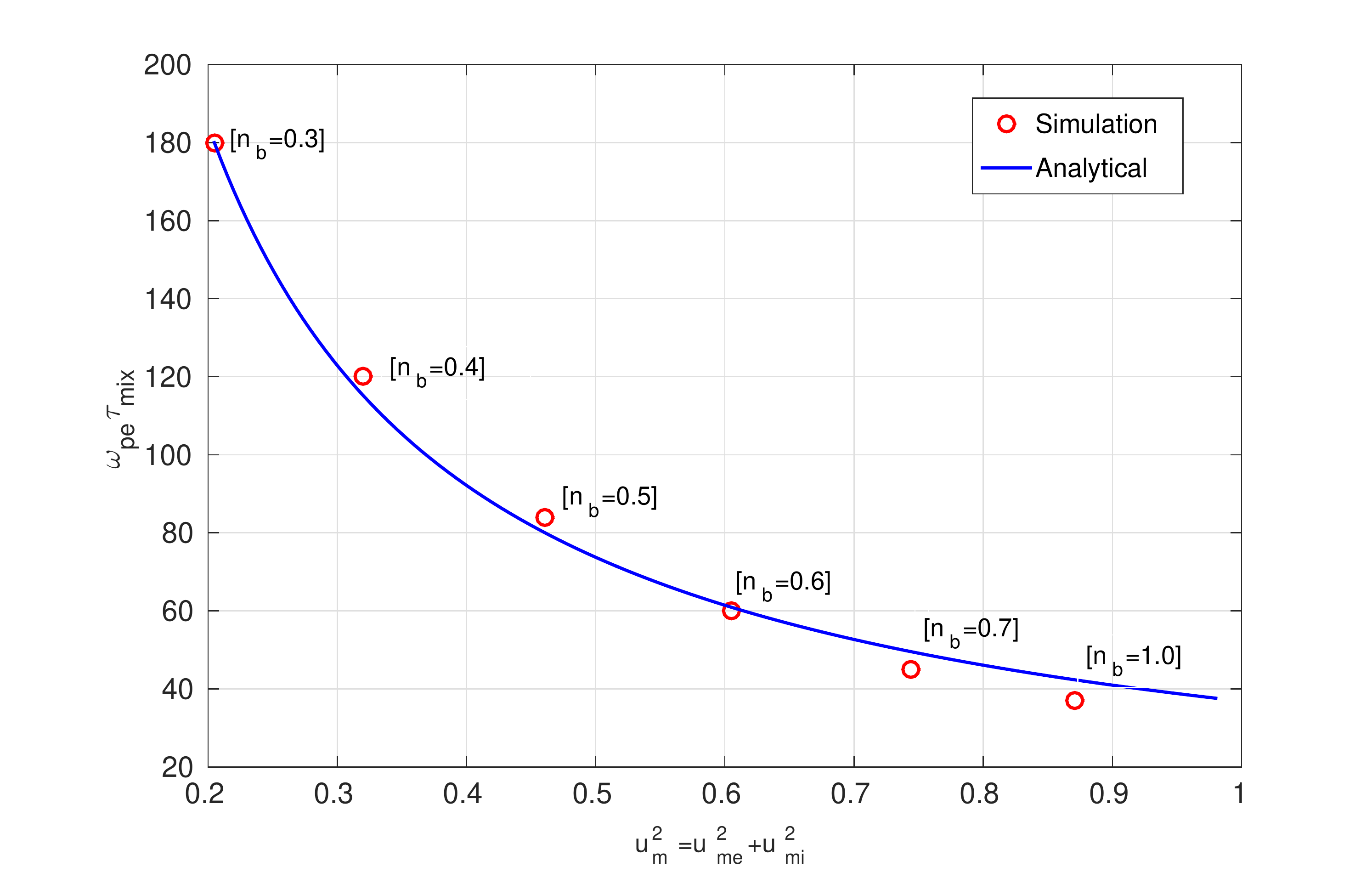}
       \caption{Plot of theoretical and numerical scaling of phase mixing time ($\omega_{pe} \tau_{mix}$)
         vs. amplitude of the waves ($u_m$).}
       \label{fig9}
\end{figure*}

\vspace{\baselineskip}
The similar observations have also been made in the present scenario but for the mobile ions. 
Due to the presence of the ion, 
though
the basic characteristics of the wave has changed from the AP mode to Khachtryan mode, but the wave breaking mechanism
is found to be similar. The beam excites a pure Khachtryan mode in which the frequency becomes
independent of space and time. As time goes, 
the original solution of Khachtryan mode excited by the beam gets perturbed by the gradual accumulation of inherent
numerical fluctuations. As a result,
the frequency of the pure Khachatryan mode becomes
space dependent and breaks via phase mixing process much below its wave breaking limit. 
Following the same methodology given in ref. \cite{arghya}
Arghya et al. \cite{arghya3} has shown the analytical 
wave breaking time for a Khachtryan mode
in terms of the wave parameters as, 
\begin{equation} 
 \tau_{mix} =\frac{2 \pi v_{ph}}{\delta^3}\left[\frac{1}{u_{me}^2 + \mu^2 u_{mi}^2}\right]
 \label{phase_mix}
\end{equation}
Where $u_{me}$, $u_{mi}$, and $\delta$ are the maximum amplitude of plasma electron velocity, 
ion velocity and the amplitude of the perturbation respectively. In our simulations for a given $\mu$, 
$u_{me}$ and $u_{mi}$ are decided by the electron beam density $n_b$. 
Typically, the value of $u_{me}$ and $u_{mi}$ increases with $n_b$ for a fixed $\mu$ (see figures (\ref{fig1})-\ref{fig2}). 
For different beam density ranging from $0.3$ to $1$ we 
have performed our simulations for $\mu=1$ and obtained the corresponding wave breaking time $\tau_{mix}$, 
$u_{me}$ and $u_{mi}$.
In Fig. (\ref{fig9}), we have plotted the $\omega_{pe}\tau_{mix}$ as a function of $u_m^2 ( =u_{me}^2 + u_{mi}^2$)
along with the 
analytical values from equation (\ref{phase_mix}). It is to be noted that the value of 
$u_{me}$ and $u_{mi}$ are equal for $\mu=1$.
For the analytical plotting in Fig. (\ref{fig9}), 
we have used the first data point ($n_b=0.3$, $u_{me}=u_{mi}=0.32$ and $\omega_{pe}\tau_{mix} =180$)
from the simulation to calculate the factor $\frac{2 \pi v_{ph}}{\delta^3}$. Using this factor,
we then draw the curve of $\omega_{pe}\tau_{mix}$ 
for different $u_m$ from equation \ref{phase_mix}. 
We see that the numerical values follow the same pattern and match with the analytical values.

\section{Conclusion}
 In summary, we have studied the excitation and breaking of relativistic electron-ion mode in a cold plasma using 1-D
fluid simulation techniques. For the excitation of the mode, we have used an external rigid homogeneous 
electron beam into the plasma. As the beam propagates through the plasma
it creates an wake wave having phase velocity equal to the velocity of the beam.
It is observed that
the excited wake wave is identical to the corresponding Khachatryan mode or
a relativistic propagating electron-ion mode in a cold plasma.
It is also found in the simulation that the numerical profile of 
the excited Khachatryan mode or wake wave gradually modifies with time and eventually 
breaks after several plasma periods. We have seen that the numerical wave breaking limit of this wave lies much 
below the analytically estimated limit given by Khachatryan et al. \cite{khacha}. 
The difference between the numerical and analytical limit has been understood 
in 
terms of phase-mixing process of this mode. We have also scaled the phase 
mixing time (or wave breaking time)
obtained from the simulations as a function of beam parameters
and found to follow the well known existing analytical scaling given by Arghya et al. \cite{arghya3} .
%
%
%
\bibliographystyle{unsrt}

%
\newpage


\end{document}